\documentclass[english]{article}
\usepackage[latin9]{inputenc}
\usepackage{geometry}
\usepackage{float}
\usepackage{mathrsfs}
\usepackage{amsmath}
\usepackage{amssymb}
\usepackage{graphicx}
\usepackage{enumerate}

\makeatletter

\floatstyle{ruled}
\newfloat{algorithm}{tbp}{loa}
\providecommand{\algorithmname}{Algorithm}
\floatname{algorithm}{\protect\algorithmname}

\@ifundefined{date}{}{\date{}}

\usepackage{float}\usepackage{mathrsfs}\usepackage{amsfonts}\usepackage{bm}\usepackage{subfigure}\usepackage{amsthm}
\usepackage{epsfig}

\usepackage[auth-sc,affil-it]{authblk}

\@ifundefined{definecolor}{\@ifundefined{definecolor}
 {\@ifundefined{definecolor}
 {\usepackage{color}}{}
}{}
}{}

\usepackage[all]{xy}

\newtheorem{prop}{Proposition}[section]

\newcounter{hypA}
\newenvironment{hypA}{\refstepcounter{hypA}\begin{itemize}
  \item[({\bf A\arabic{hypA}})]}{\end{itemize}}
\newcounter{hypB}

\usepackage{babel}

\usepackage{babel}

\begin{document}
\begin{center}

{\Large \textbf{Multilevel Monte Carlo in Approximate Bayesian Computation}}

\vspace{0.5cm}

BY  AJAY JASR$\textrm{A}^1$, SEONGIL J$\textrm{O}^2$, DAVID NOT$\textrm{T}^3$ ,
CHRISTINE SHOEMAKE$\textrm{R}^4$
\& RAUL TEMPON$\textrm{E}^5$

{\footnotesize $^{1,2,3}$Department of Statistics \& Applied Probability \& Operations Research Cluster,
National University of Singapore, Singapore, 117546, SG.}\\
{\footnotesize E-Mail:\,}\texttt{\emph{\footnotesize staja@nus.edu.sg, joseongil@gmail.com, standj@nus.edu.sg}}

{\footnotesize $^{4}$Department of Civil \& Environmental Engineering \& Operations Research Cluster,
National University of Singapore, Singapore, 119260, SG.}\\
{\footnotesize E-Mail:\,}\texttt{\emph{\footnotesize shoemaker@nus.edu.sg}}

{\footnotesize $^{5}$Center for Uncertainty Quantification
in Computational Science \& Engineering, King Abdullah University of Science and Technology, Thuwal, 23955-6900, KSA.}\\
{\footnotesize E-Mail:\,}\texttt{\emph{\footnotesize raul.tempone@kaust.edu.sa}}

\end{center}

\begin{abstract}
In the following article we consider approximate Bayesian computation (ABC) inference.
We introduce a method for numerically approximating ABC posteriors using
the multilevel Monte Carlo (MLMC). A sequential Monte Carlo version of the approach is
developed and it is shown under some assumptions that for a given level of mean square error, this method for ABC has a lower cost than i.i.d.~sampling from
the most accurate ABC approximation. Several numerical examples are given.
\end{abstract}

\section{Introduction}

In this article we are interested in inferring a particaular class of  posterior distributions in Bayesian statistics.
The scenario is when the likelihood cannot be evaluated point-wise, nor do we have access to a positive unbiased estimate of it (it is assumed we can simulate from the associated distribution, although this
is not always required).
In such a case, it is not possible to draw inference from the true posterior, even using numerical techniques such as Markov chain Monte Carlo (MCMC) or sequential Monte Carlo (SMC). The common response in Bayesian statistics, is to adopt an approximation of the posterior
using the notion of approximate Bayesian computation (ABC); see \cite{marin} for a review. ABC approximations of posteriors are based upon defining a probability distribution on an extended state-space, with
the additional random variables lying on the data-space and usually distributed according the true likelihood. The closeness of the ABC posterior distribution is controlled by a tolerance parameter $\epsilon>0$ and for some ABC approximations (but not all)
the approximation is exact as $\epsilon\rightarrow 0$. ABC has been considered in a wealth of articles and model contexts; see for instance \cite{chopin1,beaumont,jasra,jasra1,nott,pritchard,tran} for a non-exhaustive list.
In many cases of practical interest, the ABC posterior is not available exactly, and one must resort to numerical approximations, for instance using MCMC or SMC; see for instance \cite{delmoral,marin} and the references therein.

We consider using Monte Carlo to approximate expectations w.r.t.~the ABC posterior. 
Multilevel Monte Carlo \cite{giles} (see also \cite{hein:98}) methods are such that 
one sets an error threshold for a target expectation and then attains an estimator with the prescribed error 
utilizing an optimal allocation of Monte Carlo resources.  The idea assumes 
that one has a collection of approximations associated to a probability law, but the probability of interest
is intractable, even using Monte Carlo methods. For instance, it could
be a probability associated to a time-discretization of a stochastic differential equation and
the collection of approximations are finer and finer time-discretizations.
Implicitly, one is assuming that the cost associated to direct sampling of the approximations
increase with accuracy. The idea is then to rewrite the expectation w.r.t.~the most accurate
approximation and then use a telescoping sum of expecatations w.r.t.~the sequence of approximations.
Given one can appropriately sample the sequence of approximations, it can be shown for certain
models that for a given level of mean square error, MLMC has a lower cost than i.i.d.~sampling from
the most accurate approximation. See \cite{giles_acta} for a recent overview and the method is described
in more detail in Section \ref{sec:ml_abc}.

The connection between ABC and MLMC thus becomes clear; one can consider a sequence of ABC approximations
for $+\infty>\epsilon_0>\cdots>\epsilon_L>0$ and then leverage upon using the MLMC approach. There are,
however, several barriers to conducting such an approach. The first is associated to an appropriate sampling
of the sequence; the ideas of MLMC rely upon independent sampling. This issue is easily addressed,
as there exist many approaches in the literature for dependent sampling of the sequence; see for instance \cite{delmoral}.
The second and more challenging, is that the advantage of the MLMC method relies on an appropriate \emph{coupled} sampling
from the sequence of approximations. Constructing such a coupling is non-trivial for general ABC problems.
We adopt an approach which replaces coupling with importance sampling.

This paper presents an adaptation of the MLSMC method of \cite{beskos} for ABC problems. We show that, under assumptions, the use
of MLSMC is such that  for a given level of mean square error, this method for ABC has a lower cost than i.i.d.~sampling from
the most accurate ABC approximation. Several numerical examples are presented. Before our ideas are developed, we note that the MLMC method is inherently biased, in that there is approximation error, but this
error can be removed by using the ideas in \cite{rg:15} (see also \cite{graham}). This idea is cleverly utilized in \cite{tran} to perform ABC with no `$\epsilon$' error and hence is
related to the MLABC method in this paper. However, it is well-known in the ML literature that in certain contexts the variance/cost of the debiasing method
blows up, whereas, this is not the case for MLMC; see \cite{giles_acta}.

This article is structured as follows. In Section \ref{sec:ml_abc} the idea of MLMC for ABC is introduced and developed. It is
noted that in its standard form, it is not straightforward to apply in many contexts where ABC is typically used.
In Section \ref{sec:mlsmc_abc} the idea is extended to using MLSMC. Some theoretical results are considered,
showing under some assumptions that for a given level of mean square error, the MLSMC method for ABC has a lower cost than i.i.d.~sampling from
the most accurate ABC approximation. Numerical results are given in Section \ref{sec:numerics}. 
The article is concluded in Section \ref{sec:summary} with a discussion of extensions.
The appendix houses a proofs
of propositions in the article.

\section{Multilevel ABC}\label{sec:ml_abc}

\subsection{ABC Approximation}

Consider data $y\in\mathcal{Y}$, associated to finite-dimensional parameter $\theta\in\Theta\subseteq\mathbb{R}^d$.
Define the posterior:
$$
\eta_{\infty}(d\theta) \propto f(y|\theta)\pi(\theta)d\theta.
$$
We suppose that $f(y|\theta)$ is unavailable numerically, even up-to a non-negative unbiased estimator.
We consider approximate Bayesian computation (ABC). Let $E=\mathcal{Y}\times\Theta$ (with associated sigma-algebra $\mathcal{E}$) and
define for $+\infty>\epsilon_0>\cdots>\epsilon_L>0$, $x=(u,\theta)\in E$:
$$
\eta_{n}(dx) \propto K_{\epsilon_n}(y,u)f(u|\theta)\pi(\theta)d(u,\theta)
$$
where $K:\mathcal{Y}\times\mathcal{Y}\times\mathbb{R}_+\rightarrow \mathbb{R}_+$ is a user-defined non-negative function that is typically
maximized when $u=y$ and concentrates on this maximum as $\epsilon_n\rightarrow 0$. Set $Z_n=\int_EK_{\epsilon_n}(y,u)f(u|\theta)\pi(\theta)d(u,\theta)$
and $\kappa_n(x) = K_{\epsilon_n}(y,u)f(u|\theta)\pi(\theta)$.

\subsection{ML Methods}

Let $\varphi:\Theta\rightarrow\mathbb{R}_+$ with $\varphi$ bounded and measurable. Set $\eta_n(\varphi) = \int_E \varphi(\theta) \eta_n(dx)$
then we know that by the standard multilevel (ML) identity \cite{giles}:
$$
\eta_L(\varphi) = \eta_0(\varphi) + \sum_{l=1}^L [\eta_l - \eta_{l-1}](\varphi).
$$
Let $\varepsilon>0$ be given. It is known that if one can sample the coupling $(\eta_l,\eta_{l-1})$ it is possible to reduce the computational effort to achieve a given mean square error (MSE)
of $\mathcal{O}(\varepsilon^2)$, relative to i.i.d.~sampling from $\eta_L$, when approximating $\eta_{\infty}(\varphi)$. Although that is not verified for
the ABC context, we show that it is possible, with the following simple argument.

Let $(X_l,Y_l)$ be distributed from some coupling of $(\eta_l,\eta_{l-1})$, $1\leq l \leq L$.
Suppose that (call the following bullet points (A)):
\begin{itemize}
\item{$|\eta_L(\varphi)-\eta_{\infty}(\varphi)|=\mathcal{O}(\epsilon_L^{\alpha})$, for some $\alpha>0$.}
\item{$\mathbb{V}\textrm{ar}_{(\eta_l,\eta_{l-1})}[\varphi(X_l)-\varphi(Y_l)]=\mathcal{O}(\epsilon_l^{\beta})$, for some $\beta>0$.}
\item{The cost of sampling from $(\eta_l,\eta_{l-1})$ is $\mathcal{O}(\epsilon_l^{-\zeta})$, for some $\zeta>0$.}
\end{itemize}
Then supposing that $X_0$ is distributed according to $\eta_0$, one can approximate $\eta_L(\varphi)$ by
$$
\eta_0^{N_0}(\varphi) + \sum_{l=1}^L [\eta_l^{N_l} - \eta_{l-1}^{N_l}](\varphi)
$$
where for $1\leq l \leq L$ $\eta_l^{N_l}$ and $\eta_{l-1}^{N_l}$ are the empirical measures of $N_l$ independently sampled values
$(X_l^1,Y_l^1),\dots,(X_l^{N_l},Y_l^{N_l})$ from the coupling $(\eta_l,\eta_{l-1})$, independently for each $l$ and
$\eta_0^{N_0}$ is the empirical measure of $N_0$ independent samples from $\eta_0$ (independent of all other random variables). 
Then the MSE is
$$
\mathbb{E}[(\eta_0^{N_0}(\varphi) + \sum_{l=1}^L [\eta_l^{N_l} - \eta_{l-1}^{N_l}](\varphi) - \eta_{\infty}(\varphi))^2] 
=
$$
$$
|\eta_L(\varphi)-\eta_{\infty}(\varphi)|^2 + \frac{1}{N_0}\mathbb{V}\textrm{ar}_{\eta_0}[\varphi(X_0)]
+ \sum_{l=1}^L \frac{1}{N_l}\mathbb{V}\textrm{ar}_{(\eta_l,\eta_{l-1})}[\varphi(X_l)-\varphi(Y_l)].
$$
Setting $\epsilon_l = M^{-l}$ for some fixed integer $M>1$ if we want the MSE to 
be $\mathcal{O}(\varepsilon^2)$ we can make the bias and variance this order. So we want
$$
\epsilon_L^{2\alpha} = M^{-2L} = \varepsilon^2
$$
so $L=\mathcal{O}(|\log(\varepsilon)|)$.
Now we require
$$
\sum_{l=0}^L \frac{\epsilon_l^{\beta}}{N_l} = \mathcal{O}(\varepsilon^2)
$$
and at the same time, we seek to minimize the cost of doing so $\sum_{l=0}^L N_l \epsilon_l^{-\zeta}$.
This constrained optimization problem is easily solved with Lagrange multipliers (e.g.~\cite{giles}) yielding
that
$$
N_l = \varepsilon^{-2}\epsilon_l^{(\beta+\zeta)/2} K_L
$$
where $K_L =\sum_{l=0}^L\epsilon_l^{(\beta-\zeta)/2}$. Under this choice 
$$
\sum_{l=0}^L \frac{\epsilon_l^{\beta}}{N_l} = \varepsilon^{2} K_{L}^{-1} \sum_{l=0}^L\epsilon_l^{(\beta-\zeta)/2} = \mathcal{O}(\varepsilon^2).
$$
This yields a cost of $\varepsilon^{-2}K_L^2$.
The cost of i.i.d.~sampling from $\eta_L$ to achieve a MSE of $\mathcal{O}(\varepsilon^2)$ is
$\varepsilon^{-2}\epsilon_L^{-\zeta}$. If $\beta\geq\zeta$ then the MLMC method certainly outperforms
i.i.d.~sampling from $\eta_L$. The worst scenario is when $\beta<\zeta$.  In this case 
it is sufficient to set $K_{L}=\epsilon_L^{(\beta-\zeta)/2}$ to make the variance 
$\mathcal{O}(\varepsilon^2)$, and then 
the number of samples on the finest level is given by $N_L = \epsilon_L^{\beta-2\alpha}$
whereas the total algorithmic cost is
$\mathcal{O}(\varepsilon^{-(\zeta/\alpha + \delta)})$, where $\delta = 2-\beta/\alpha \geq 0$.
In this case, one can choose the largest value for the bias, $\alpha = \beta/2$, so that $N_L=1$ and the total cost, $\mathcal{O}(\varepsilon^{-\zeta/\alpha})$,  is dominated by this single
sample. We remark that when debiasing this procedure and $\beta<\zeta$ using \cite{rg:15} the variance/cost blows up.

The issue with this construction, ignoring verifying (A), is that 
in an ABC context, it is challenging to construct the coupling and even if one can, 
seldom can one achieve i.i.d.~sampling from the couples.

\section{Multilevel Sequential Monte Carlo for ABC}\label{sec:mlsmc_abc}

\subsection{Approach}

The approach in \cite{beskos} is to by-pass the issue of coupling, by using importance sampling and then to use sequential Monte Carlo (SMC)
\cite{delm:06}
samplers to provide the appropriate simulation. Set $G_n(x) = \kappa_{n+1}(x)/\kappa_{n}(x)$. Then \cite{beskos} show that
\begin{equation}
\eta_L(\varphi) = \frac{Z_0}{Z_1}\eta_0(G_0\varphi) + \sum_{l=2}^L \eta_{l-1}\Big(\big(\frac{Z_{l-1}}{Z_l}G_{l-1}-1\big)\varphi\Big)\label{eq:ml_approx}.
\end{equation}
\cite{beskos} show how such an identity can be approximated as we now describe.

It is remarked that much of the below information is in \cite{beskos} and is necessarily recalled here.
We will apply an SMC sampler to obtain 
a collection of samples (particles) that sequentially approximate $\eta_0, \eta_1,\ldots, \eta_{L-1}$. 
We consider the case when one initializes the population of particles by sampling  i.i.d.~from $\eta_0$, then at every step  resamples and applies a MCMC kernel to mutate the particles.
We denote by $(X_{0}^{1:N_0},\dots,X_{L-1}^{1:N_{L-1}})$, with $+\infty > N_0\geq N_1\geq \cdots  N_{L-1}\geq 1$, the samples after mutation; one resamples $X_l^{1:N_l}$ according to the weights $G_{l}(X_l^i) = 
(\kappa_{l+1}/\kappa_l)(X_l^{i})$, for indices $l\in\{0,\dots,L-1\}$.
Let $\{M_l\}_{1\leq l\leq L-1}$ denote a sequence of MCMC kernels, with the property $\eta_{l}M_l = \eta_l$.
These kernels are used at stages $1,\dots,L-1$ of the SMC sampler.
For $\varphi:E\rightarrow\mathbb{R}$, $l\in\{1,\dots,L\}$, we have the following estimator 
of $\mathbb{E}_{\eta_{l-1}}[\varphi(X)]$:
$$
\eta_{l-1}^{N_{l-1}}(\varphi) = \frac{1}{N_{l-1}}\sum_{i=1}^{N_{l-1}}\varphi(X_{l-1}^i)\ . 
$$
We define
$$
\eta_{l-1}^{N_{l-1}}(G_{l-1}M_l(dx_l)) = \frac{1}{N_{l-1}}\sum_{i=1}^{N_{l-1}}G_{l-1}(X_{l-1}^i) M_l(X_{l-1}^i,dx_l)\ .
$$
The joint probability distribution for the SMC algorithm is 
$$
\prod_{i=1}^{N_0} \eta_0(dx_0^i) \prod_{l=1}^{L-1} \prod_{i=1}^{N_l} \frac{\eta_{l-1}^{N_{l-1}}(G_{l-1}M_l(dx_l^i))}{\eta_{l-1}^{N_{l-1}}(G_{l-1})}\ .
$$
If one considers one more step in the above procedure, that would deliver samples 
$\{X_L^i\}_{i=1}^{N_L}$, a standard SMC sampler estimate of the quantity of interest in (\ref{eq:ml_approx})
is $\eta_L^{N_L}(g)$; the earlier samples are discarded. 
An SMC approximation of \eqref{eq:ml_approx}
$$
\widehat{Y} =
\sum_{l=2}^{L}\Big\{\frac{\eta_{l-1}^{N_{l-1}}(\varphi G_{l-1})}{\eta_{l-1}^{N_{l-1}}(G_{l-1})} - \eta_{l-1}^{N_{l-1}}(\varphi)\Big\}
+\frac{\eta_{0}^{N_{0}}(\varphi G_{0})}{\eta_{0}^{N_{0}}(G_{0})}.
$$

\cite[Theorem 1]{beskos} shows that the MSE of the MLSMC method is upper-bounded by 
$$
|\eta_L(\varphi)-\eta_{\infty}(\varphi)|^2 + \frac{C}{N_0} + C\sum_{l=2}^L \frac{1}{N_{l-1}}\Big\|\frac{Z_{l-1}}{Z_l}G_{l-1}-1\Big\|_{\infty}^2
+
$$
\begin{equation}\label{eq:mss_smc_bound}
\sum_{2\leq l <q\leq L}\bigg\{ \Big\|\frac{Z_{l-1}}{Z_l}G_{l-1}-1\Big\|_{\infty}
\Big\|\frac{Z_{q-1}}{Z_q}G_{q-1}-1\Big\|_{\infty}
\Big(
\frac{\kappa^{q-1}}{N_{l-1}} + \frac{1}{N_{l-1}^{1/2}N_{q-1}}
\Big)\bigg\}
\end{equation}
where $\|\cdot\|_{\infty}$ is the supremum norm and $C<+\infty$, $\kappa\in(0,1)$ are constants that do not depend upon $l,q$.
\cite{beskos} use the following assumptions, which we will consider in the analysis of MLSMC in the ABC context.
Note that these assumptions have been weakened in \cite{djl_md}.
\begin{hypA}
\label{hyp:A}
There exist $0<\underline{C}<\overline{C}<+\infty$ such that
\begin{eqnarray*}
\sup_{l \geq 1} 
\sup_{u\in E} G_l (u) & \leq & \overline{C}\ ;\\
\inf_{l \geq 1} 
\inf_{u\in E} G_l (u) & \geq & \underline{C}\ .
\end{eqnarray*}
\end{hypA}

\begin{hypA}
\label{hyp:B}
There exists a $\rho\in(0,1)$ such that for any $l\ge 1$, $(x,z)\in E^2$, $A\in\mathcal{E}$:
$$
\int_A M_l(x,dx') \geq \rho \int_A M_l(z,dz')\ .
$$
\end{hypA}

One can see, in (A\ref{hyp:B}), that the MCMC kernel must mix uniformly well w.r.t.~the level indicator.
If the MCMC kernel cost is $\mathcal{O}(1)$ (i.e.~independent of $\epsilon$) then one can iterate to (e.g.) $\mathcal{O}(\epsilon_l^{-\zeta})$ at a given level $l$.
That is, as one expects the complexity of the posterior to increase as $\epsilon$ falls, one must put in more effort to efficiently sample the posterior and achieve a uniform mixing rate.
In other situations, the cost of the MCMC step may directly depend upon $\epsilon_l$, in order for the mixing rate to be uniform in $l$.



\subsection{Some Analysis}

In order to understand the utility of applying MLSMC for ABC, we must understand the MSE and in particular, terms such
as
$$
\Big\|\frac{Z_{l-1}}{Z_l}G_{l-1}-1\Big\|_{\infty}.
$$
We show that under fairly general assumptions, that this expression can be controlled in terms of $\epsilon_{l-1}$.
It is supposed that $\Theta$ and $\mathcal{Y}\subset\mathbb{R}^n$ (for some $n\geq 1$ be given) are compact and we take:
\begin{equation}\label{eq:kernel}
 K_{\epsilon_l}(y,u) = \prod_{i=1}^n \frac{1}{1+ \big(\frac{y_i-u_i}{\epsilon_l}\big)^2}.
\end{equation}
This is a quite general context, as we do not assume anything more about $f(u|\theta)$ and $\pi(\theta)$.
It is supposed that $\epsilon_{l-1}/\epsilon_l = \mathcal{O}(1)$, which is reasonable (e.g.~$\epsilon_l=M^{-l}$).

In this scenario, it is straightforward to show that for any $x\in E$
$$
\underline{C}\leq G_l(x) \leq \overline{C}
$$
for any fixed $0\leq l \leq L_1$ where $\underline{C},\overline{C}$ do not depend on $l$;
this verifies (A\ref{hyp:A}).
We have the following result, the proof of which, is in the appendix:

\begin{prop}\label{prop:norm_control}
Let $n\geq 1$ be given. Then there exists a $C>0$ such that for any $1\leq l \leq L$: 
$$
\Big\|\frac{Z_{l-1}}{Z_l}G_{l-1}-1\Big\|_{\infty} \leq C \epsilon_{l-1}^2.
$$ 
\end{prop}

Suppose that the cost of one independent sample from $\eta_l$ is $\epsilon_l^{-\zeta}$ and that our MCMC kernel also costs the same.
Given $\varepsilon>0$, and supposing the bias of $\mathcal{O}(\epsilon_L^{\alpha})$, $\epsilon_l=M^{-l}$  $L=\mathcal{O}(|\log(\varepsilon)|)$  
the procedure for finding the optimal $N_{0:L-1}$ to minimize the cost $\sum_{l=0}^{L-1} \epsilon_l^{-\zeta}N_l$ so that the variance is $\mathcal{O}(\varepsilon^2)$
is as in \cite{beskos}. The idea there is to just consider the term
$$
\sum_{l=1}^L \frac{1}{N_l}\Big\|\frac{Z_{l-1}}{Z_l}G_{l-1}-1\Big\|_{\infty}^2
$$
in the variance part of the bound \eqref{eq:mss_smc_bound}. The constrained optimization is then as in \cite{giles}.
We then check that the additional term in \eqref{eq:mss_smc_bound} is $\mathcal{O}(\varepsilon^2)$ or smaller.
Therefore, setting, $N_l = \varepsilon^{-2}\epsilon_l^{(4+\zeta)/2} K_L$, the variance part of
\eqref{eq:mss_smc_bound} is
$$
\varepsilon^{2} K_L^{-1} \sum_{l=1}^L \epsilon_{l-1}^{(4-\zeta)/2} +
\sum_{2\leq l <q\leq L}\epsilon_{l-1}^2\epsilon_{q-1}^2
\Big[\frac{\varepsilon^2\kappa^{q-1}}{K_L\epsilon_{l-1}^{(4+\zeta)/2}}
+ \frac{\varepsilon^3}{K_L^{3/2}\epsilon_{l-1}^{(4+\zeta)/4}\epsilon_{q-1}^{(4+\zeta)/2}}
\Big].
$$
As shown in \cite[Section 3.3]{beskos} if $\zeta\leq 2\alpha$ then the additional term is $\mathcal{O}(\varepsilon^2)$.
So therefore, the conclusion is as in Section \ref{sec:ml_abc} (the cost is the same as discussed there): for a given level of MSE, the MLSMC method for ABC has a lower cost than i.i.d.~sampling from $\eta_L$. The main issue is to determine the bias, which often needs to be model specific;
we give an example where this is possible.

\subsection{Example}
\label{sec:example}
We consider a state-space model. Let $y=(v_{0:n})\in\mathcal{Y}=\mathsf{V}^n$
and $\theta=(w_{0:n})\in\Theta=\mathsf{W}^n$, where we suppose $\mathsf{V},\mathsf{W}$
are compact subsets of a power of the real-line. In a state-space model, we can write:
$$
p(v_{0:n},w_{0:n}) = \mu(w_0)g(v_0|w_0)\prod_{i=1}^n g(v_i|w_i)h(w_i|w_{i-1})
$$
where $p(v_{0:n},w_{0:n})$ is the joint density of the random variables $(y,\theta)$,
$\mu$ is a probability density on $\mathsf{W}$, for any $w\in\mathsf{W}$, $g(\cdot|w)$
(resp.~$h(\cdot|w)$) is a probability density on $\mathsf{V}$ (resp~$\mathsf{W}$).

We are interested in the posterior:
$$
\eta_{\infty}(d\theta) \propto p(v_{0:n},w_{0:n}) dw_{0:n}.
$$
If $g$ and $h$ are intractable in some way, but can be sampled (although this is not a requirement - see \cite{jasra} and the references therein), then an ABC approximation is:
\begin{equation}
\label{eq:abc_approx}
\eta_{l}(dx) \propto K_{\epsilon_l}(y-u)p(u_{0:n},w_{0:n}) d(u_{0:n},w_{0:n})
\end{equation}
with $u=u_{0:n}\in\mathcal{Y}$. Let $\vartheta:\mathsf{W}\rightarrow\mathbb{R}$ be bounded and measurable and $\varphi(\theta)=
\sum_{p=0}^n\vartheta(w_p)$. Then, under the assumptions in \cite{martin}
$$
|[\eta_{l}-\eta_{\infty}](\varphi)| \leq C\|\vartheta\|_{\infty}\epsilon_l
$$
where $C$ depends linearly on $n$, so that the bias assumption of (A) is satisfied with $\alpha=1$ for additive functionals.

Suppose one uses a single site Gibbs sampler as the MCMC kernel.
Let $1\leq l \leq L$ and for a vector $z_{0:n}$ set $z_{-i}$ be all the elements except the $i^{th}$, $i\in\{0,\dots,n\}$,
then for each $i\in\{1,\dots,n\}$
sampling is performed from
$$
\eta_l(d(v_i,w_i)|v_{-i},w_{-i}) \propto \frac{1}{1+ \big(\frac{y_i-v_i}{\epsilon_l}\big)^2} h(w_{i+1}|w_{i}) g(v_i|w_i)h(w_i|w_{i-1}) d(v_i,w_i)
$$
with the case $i=0$ 
$$
\eta_l(d(v_0,w_0)|v_{-0},w_{-0}) \propto \frac{1}{1+ \big(\frac{y_0-v_0}{\epsilon_l}\big)^2} h(w_{1}|w_{0}) g(v_0|w_0)\mu(w_0) d(v_0,w_0).
$$
It is simple to show that (A\ref{hyp:B}) is satisfied (the constants depend upon $n$). That is, that writing the density of the kernels as $M_l$
it can be shown that
$$
\frac{M_l(x,x')}{M_l(z,z')} \geq C
$$
for any fixed $x,x',z,z'$ and $C$ is independent of $l$. Moreover, if one samples from the full conditionals using rejection sampling with proposal
when $i\in\{1,\dots,n\}$ (resp.~$i=0$) $g(v_i|w_i)h(w_i|w_{i-1}) d(v_i,w_i)$ (resp.~$g(v_0|w_0)\mu(w_0) d(v_0,w_0)$), we have the following result, whose proof is in the appendix:
\begin{prop}\label{prop:cost}
The expected cost of one iteration of the above Gibbs sampler is $\mathcal{O}(n\epsilon_{l}^{-1})$.
\end{prop}
In this example for a given level of MSE, the MLSMC method for ABC has a lower cost than i.i.d.~sampling from $\eta_L$ as the associated (exact independent) rejection sampling
cost is $\mathcal{O}(\epsilon_{L}^{-n})$ and the cost of sampling $\eta_0$ is $\mathcal{O}(1)$.


\section{Numerical Examples}\label{sec:numerics}

\subsection{Linear Gaussian State-Space Model}

We now consider some simulations in the context of the example in Section \ref{sec:example}.
In this case, $\mathsf{V}=\mathsf{W}=\mathbb{R}$ and we take:
\begin{eqnarray*}
V_i|W_i=w_i & \sim & \mathcal{N}(w_i,\sigma^2_v) \quad i\geq 0\\
W_i |W_{i-1}=w_{i-i}  & \sim & \mathcal{N}(w_{i-1},\sigma^2_w) \quad i\geq 1
\end{eqnarray*}
where $\mathcal{N}(\mu,\sigma^2)$ is a Gaussian distribution of mean $\mu$ and variance $\sigma^2$,
with $\mu(w_0) \sim \mathcal{N}(0,\sigma^2_w)$ and both $\sigma^2_w,\sigma_v^2>0$ given constants.
The ABC approximation is taken as in Section \ref{sec:example} equation \eqref{eq:abc_approx} with kernel as in \eqref{eq:kernel}.
In this scenario, there is of course no reason to use ABC methods, however, one can compute the exact value
of (for instance) $\mathbb{E}[W_i|v_{0:i}]$ exactly, which allows us to compute accurate MSEs. The data are generated from the model.

\begin{figure}\centering
  \includegraphics[height=6cm,width=0.7\textwidth]{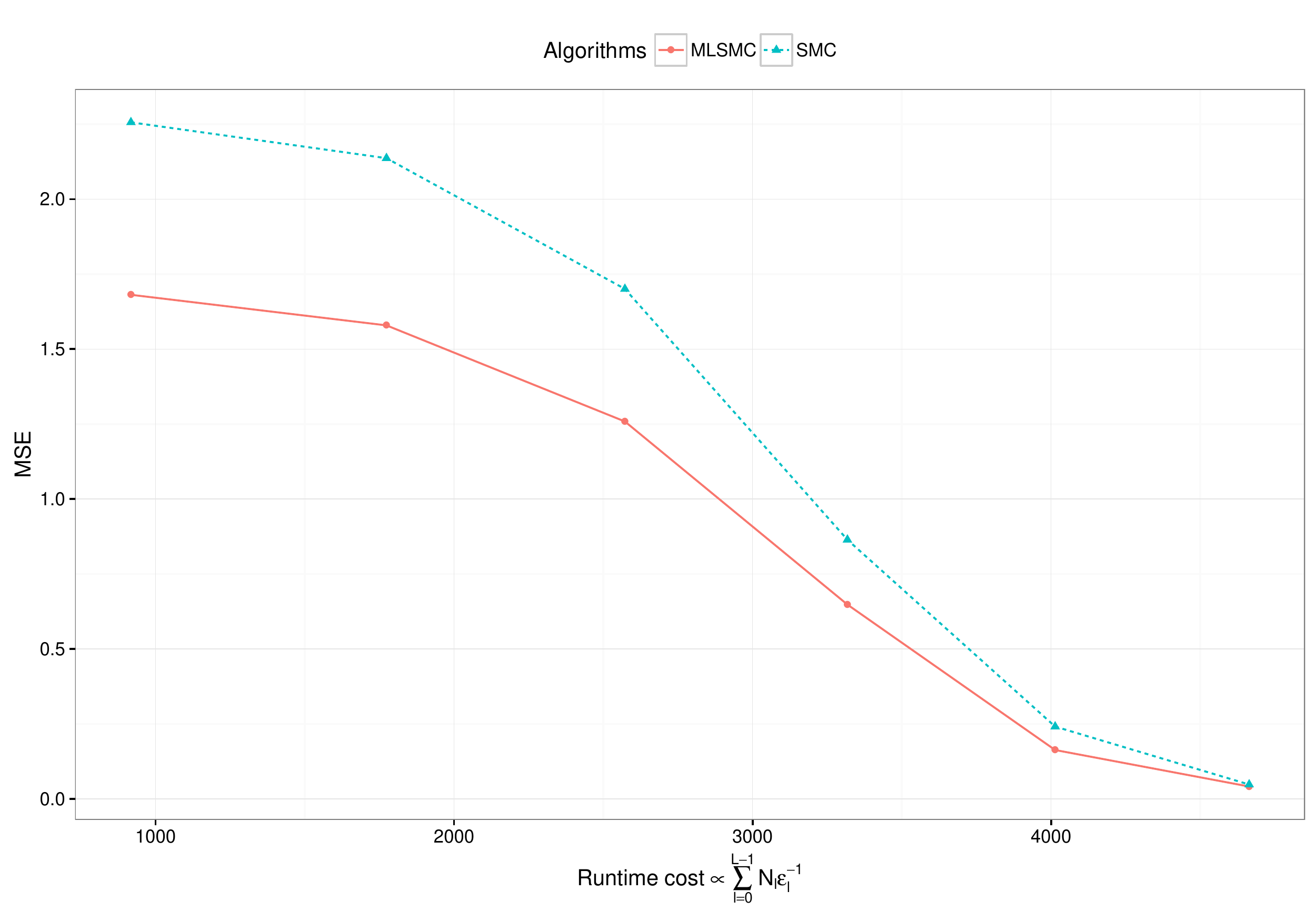}
  \caption{Mean Square Error against Cost. This is for the linear Gaussian state-space model, $n=10$.}
  \label{fig:T10}
\end{figure}

\begin{figure}\centering
  \includegraphics[height=6cm,width=0.7\textwidth]{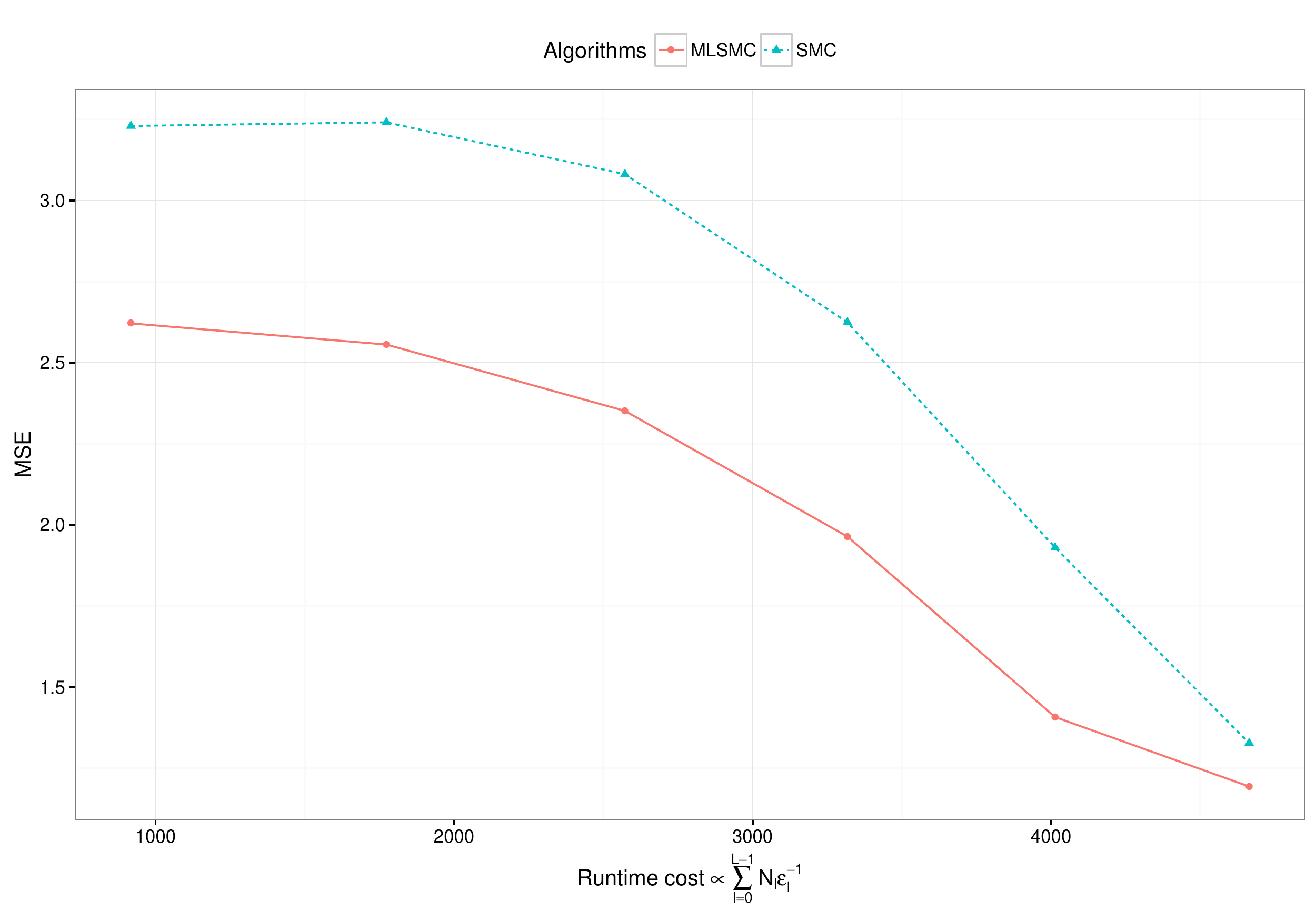}
  \caption{Error against Cost. This is for the linear Gaussian state-space model, $n=25$.}
  \label{fig:T25}
\end{figure}

We will compare the MLSMC method of this article to an SMC sampler (such as in \cite{delmoral} with no adaptation) that has approximately the same computational
cost. By SMC sampler, we simply mean that the number of samples used at each time step of the SMC algorithm is the same
and only the samples which approximate $\eta_L$ are used to estimate expectations w.r.t.~this distribution.
We will consider the estimate of 
$$
\int_E w_n \eta_L(dx).
$$
As noted above, if $\epsilon_L=0$ then one knows this value exactly. We set $\epsilon_l=C2^{-l}$ and $L=5$
and consider the cases $n\in\{10,25\}$. The MCMC kernel adopted is a single-site Metropolis-Hastings kernel with
proposals as in Section \ref{sec:example}. That is,  for $i\in\{1,\dots,n\}$ (resp.~$i=0$) $g(v_i|w_i)h(w_i|w_{i-1}) d(v_i,w_i)$ (resp.~$g(v_0|w_0)\mu(w_0) d(v_0,w_0)$). In the MLSMC sampler, we set $N_l = \varepsilon^{-2}\epsilon_l^{(4+1)/2} K_L$
with $\varepsilon$ variable across examples - 6 different values are run. The SMC sampler is run so that the computational
run-time is almost the same. We repeat our simulations 10 times. The results are given in Figures \ref{fig:T10}-\ref{fig:T25}.

Figures \ref{fig:T10}-\ref{fig:T25} show that for the scenario under study, the MLSMC sampler out-performs the standard SMC
sampler approach, as is also shown in \cite{beskos}. Even though some of the mathematical assumptions that are made in \cite[Theorem 1]{beskos}
are violated, the predicted improvement at almost no extra coding effort is seen in practice. 

\subsection{Intractable State-Space Model}

We consider the stochastic volatility model (SVM) given by
\begin{eqnarray*}
	V_i \mid W_i = w_i & \sim & {\mathcal St}\left(0, \exp(w_i/2), s_2, s_3\right), ~~ i \ge 1, \\
	W_i \mid W_{i-1} = w_{i-1} & \sim & {\mathcal N}\left(\alpha + \beta \left(w_{i-1} - \alpha\right), \sigma_w^2\right), ~~ i \ge 2, \\
	W_1 & \sim & {\mathcal N}\left(\alpha, \frac{\sigma^2_w}{1 - \beta^2}\right), 
\end{eqnarray*}
where $V_i$ are the mean-corrected returns and ${\mathcal St}(s_0, s_1, s_2, s_3)$ denotes a stable distribution with location parameter $s_0$, scale parameter $s_1$, asymmetry parameter $s_2$ and skewness parameter $s_3$. We set $s_2 = 1.75$ and $s_3 = 1$ as in \cite{jasra}. To guarantee stationarity of the latent log-volatility ${W_i}$, we assume that $|\beta| < 1$. 

We assign priors $\alpha \sim {\mathcal{N}}\left(0, 100\right)$, $\beta \sim {\mathcal N}\left(0, 10\right)$ on $(-1, 1)$ and $\sigma^2_w \sim {\mathcal{IG}}\left(2, 1/100\right)$. Note $\mathcal{IG}(2, 1/100)$ is an inverse gamma distribution with mean $1/100$ and infinite variance. 
The ABC approximation is taken as in Section \ref{sec:example} equation \eqref{eq:abc_approx} with kernel as in \eqref{eq:kernel}.

We use the daily index of the S\&P 500 index between 1 January 2011--2014 February 2013 (533 data points).
The dataset can be obtained from {\tt http://ichart.finance.yahoo.com}.
We first estimate the value of $\eta_\infty$ using the MLSMC algorithm with $L = 7$, and then we compare the MLSMC sampler with the SMC sampler with $L = 5$ as in examples for the linear Gaussian state-space model. We again set $\epsilon_l = C2^{-l}$ and $N_l = \varepsilon^{-2}\epsilon_l^{(4+1)/2}K_L$.
For the MCMC kernel, we adapt a single-site Metropolis-Hastings kernel with proposals as in \cite{Celeux06}.

\begin{figure}\centering
  \includegraphics[height=6cm,width=0.7\textwidth]{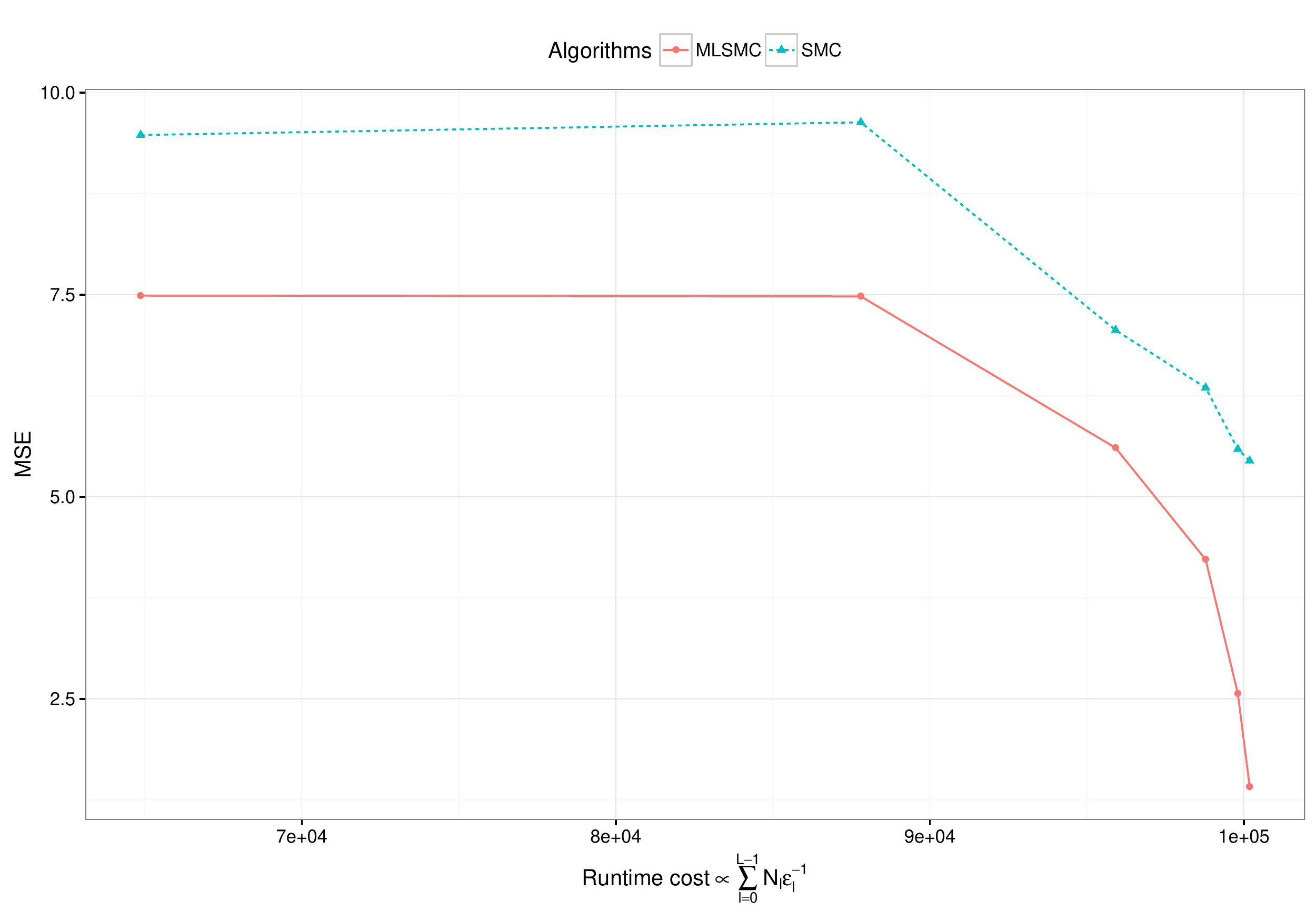}
  \caption{Error against Cost. This is for the SVM.}
  \label{fig:svm}
\end{figure}

The results, when estimating the same functional as for the linear Gaussian model, can be found in Figure \ref{fig:svm}. The Figure shows as for the previous example that the MLSMC procedure is out-performing using SMC, in the sense that the MSE for a given cost is lower for the former approach.

\section{Summary}\label{sec:summary}

In this article we have considered the development of the MLMC method in the context of ABC.
Several extensions of this work are possible.
The first is that, it is well-known that the sequence of $\epsilon$ can be set on the fly, using an adaptive SMC method.
It is then of interest to see if MLSMC has a benefit from a theoretical perspective (see e.g.~\cite{beskos1} for an analysis of adaptive SMC).
The second is the consideration of the possible improvement of MLSMC when the summary statistics of ABC are not sufficient, as they
have been in this paper.

\subsubsection*{Acknowledgements}
AJ, SJ, DN \& CS were all supported by grant number R-069-000-074-646, Operations research cluster funding, NUS.

\appendix

\section{Proof of Proposition \ref{prop:norm_control}}

\begin{proof}
We give the proof in the case $n=1$; the general case is the same, except with some minor complications in notations.
We have
\begin{equation}\label{eq:master}
\frac{Z_{l-1}}{Z_l}G_{l-1}(x) - 1 = 
\frac{Z_{l-1}}{Z_l}\Big(G_{l-1}(x) - \frac{\epsilon_l^2}{\epsilon_{l-1}^2}\Big) + 
\frac{\epsilon_l^2}{\epsilon_{l-1}^2}\frac{Z_{l-1}}{Z_l} - 1.
\end{equation}
We will deal with the two expressions on the R.H.S.~of \eqref{eq:master} separately.
Throughout $C$ is a constant that does not depend on a level index $l$ but whose value
may change upon appearance.

\textbf{First Term on the R.H.S.~of \eqref{eq:master}}

We will show that 
$$
\Big\|\Big(G_{l-1} - \frac{\epsilon_l^2}{\epsilon_{l-1}^2}\Big) \Big\|_{\infty} \leq C \epsilon_{l-1}^2
$$
and that $\frac{Z_{l-1}}{Z_l}\leq C$. We start with the first task. We have
$$
G_{l-1}(x) - \frac{\epsilon_l^2}{\epsilon_{l-1}^2} = \frac{\epsilon_l^2}{\epsilon_{l-1}^2}\Big[
\frac{\epsilon_{l-1}^2+c}{\epsilon_{l}^2+c}
-1
\Big].
$$
where we have set $c=(y-u)^2$. Then elementary calculations yield
$$
G_{l-1}(x) - \frac{\epsilon_l^2}{\epsilon_{l-1}^2} = 
\frac{\epsilon_{l}^2}{\epsilon_{l}^2+c}\Big(1-\frac{\epsilon_{l}^2}{\epsilon_{l-1}^2}\Big).
$$
Now as
$$
\frac{\epsilon_{l}^2}{\epsilon_{l-1}^2}\leq 1 
$$
and as $c\geq C$
$$
\frac{1}{\epsilon_{l}^2+c} \leq \frac{1}{c} \leq C
$$
we have 
\begin{equation}\label{eq:pt1}
\Big\|\Big(G_{l-1} - \frac{\epsilon_l^2}{\epsilon_{l-1}^2}\Big) \Big\|_{\infty} \leq C \epsilon_{l}^2
\leq C \epsilon_{l-1}^2.
\end{equation}
Now 
$$
\frac{Z_{l-1}}{Z_l} = \int_E \frac{1}{1+\frac{c}{\epsilon_{l-1}^2}} f(u|\theta)\pi(\theta)d(\theta,u) \Big(\int_E \frac{1}{1+\frac{c}{\epsilon_{l}^2}} f(u|\theta)\pi(\theta)d(\theta,u)\Big)^{-1}.
$$
Now
$$
\frac{1}{1+\frac{c}{\epsilon_{l-1}^2}} \leq C \epsilon_{l-1}^2
$$
so that 
$$
Z_{l-1} \leq C \epsilon_{l-1}^2.
$$
We now will show that $\epsilon_{l-1}^{-2} Z_l$ is lower bounded uniformly in $l$ which will show that $\frac{Z_{l-1}}{Z_l}\leq C$.
$$
\epsilon_{l-1}^{-2} Z_l = \int_E \frac{1}{\epsilon_{l-1}^{2}+\frac{c\epsilon_{l-1}^{2}}{\epsilon_{l}^2}} f(u|\theta)\pi(\theta)d(\theta,u).
$$
Then
$$
\frac{1}{\epsilon_{l-1}^{2}+\frac{c\epsilon_{l-1}^{2}}{\epsilon_{l}^2}} \geq \frac{1}{1 + C}
$$
as $\epsilon_{l-1}^{2}\leq 1$ and $\frac{c\epsilon_{l-1}^{2}}{\epsilon_{l}^2}\leq C$. So we have that $\epsilon_{l-1}^{-2} Z_l\geq C$
and
\begin{equation}\label{eq:pt2}
\frac{Z_{l-1}}{Z_l}\leq C.
\end{equation}
Combining \eqref{eq:pt1} and \eqref{eq:pt2} yields
\begin{equation}\label{eq:pt3}
\Big\|\Big(G_{l-1} - \frac{\epsilon_l^2}{\epsilon_{l-1}^2}\Big) \Big\|_{\infty} \leq C \epsilon_{l-1}^2.
\end{equation}

\textbf{Second Term on the R.H.S.~of \eqref{eq:master}}

Clearly
\begin{equation}\label{eq:simple_decomp}
\frac{\epsilon_l^2}{\epsilon_{l-1}^2}\frac{Z_{l-1}}{Z_l} - 1 = \frac{\epsilon_l^2 Z_{l-1} - \epsilon_{l-1}^2 Z_{l}}{\epsilon_{l-1}^2 Z_{l}}.
\end{equation}
We first deal with the numerator on the R.H.S.:
\begin{eqnarray*}
\epsilon_l^2 Z_{l-1} - \epsilon_{l-1}^2 Z_{l} & = & \int_E 
\Big[\frac{\epsilon_{l-1}^2\epsilon_{l}^2}{c+\epsilon_{l-1}^2 } -
\frac{\epsilon_{l-1}^2\epsilon_{l}^2}{c+\epsilon_{l}^2 }\Big]
 f(u|\theta)\pi(\theta)d(\theta,u)\\ & = &
\int_E 
\Big[\frac{\epsilon_{l-1}^2\epsilon_{l}^2(\epsilon_{l}^2-\epsilon_{l-1}^2)}{(c+\epsilon_{l-1}^2)(c+\epsilon_{l}^2)}\Big]
 f(u|\theta)\pi(\theta)d(\theta,u).
\end{eqnarray*}
As 
$$
\frac{1}{(c+\epsilon_{l-1}^2)(c+\epsilon_{l}^2)} \leq C, \quad\quad \epsilon_{l}^2-\epsilon_{l-1}^2 \leq \epsilon_{l}^2 \leq \epsilon_{l-1}^2
$$
we have
$$
|\epsilon_l^2 Z_{l-1} - \epsilon_{l-1}^2 Z_{l}| \leq C\epsilon_{l-1}^4\epsilon_{l}^2.
$$
Therefore one has
$$
\Big|\frac{\epsilon_l^2 Z_{l-1} - \epsilon_{l-1}^2 Z_{l}}{\epsilon_{l-1}^2 Z_{l}}\Big| \leq  C\frac{\epsilon_{l-1}^2}{\epsilon_{l}^{-2} Z_l}
$$
Using almost the same calculation as for showing $\epsilon_{l-1}^{-2} Z_l\geq C$, we have
$\epsilon_{l}^{-2} Z_l\geq C$
and so
\begin{equation}\label{eq:pt4}
\Big|\frac{\epsilon_l^2}{\epsilon_{l-1}^2}\frac{Z_{l-1}}{Z_l} - 1\Big| \leq C\epsilon_{l-1}^2.
\end{equation}

Returning to \eqref{eq:master} and noting \eqref{eq:simple_decomp}, one apply the triangular inequality
and combine \eqref{eq:pt2} and \eqref{eq:pt4} to complete the proof.
\end{proof}

\section{Proof of Proposition \ref{prop:cost}}

\begin{proof}
We will show that the expected cost of sampling a given full-conditional is $\mathcal{O}(\epsilon_l^{-1})$. 
Throughout $C$ is a constant that does not depend on a level index $l$ nor $i$ but whose value
may change upon appearance.

It is easily shown that
$$
h(w_{i+1}|w_{i}) \frac{1}{1+ \big(\frac{y_i-v_i}{\epsilon_l}\big)^2}  \leq C=C^*
$$
and thus that the probability of accepting, in the rejection scheme 
is
$$
(C^*)^{-1}\int_{\mathsf{V}\times\mathsf{W}} \frac{1}{1+ \big(\frac{y_i-v_i}{\epsilon_l}\big)^2} h(w_{i+1}|w_{i}) g(v_i|w_i)h(w_i|w_{i-1}) d(v_i,w_i).
$$
The expected number of simulations is then the inverse. Clearly
$$
\frac{1}{1+ \big(\frac{y_i-v_i}{\epsilon_l}\big)^2} \geq C\epsilon_l
$$
and as $\mathsf{V}\times\mathsf{W}$ is compact $h(w_{i+1}|w_{i}) g(v_i|w_i)h(w_i|w_{i-1})\geq C$ so that the expected number of simulations to sample the full conditional is at most
$$
C\epsilon_l^{-1}
$$
which completes the proof.
\end{proof}

\end{document}